\documentclass[preprint,showpacs,preprintnumbers,amsmath,amssymb,prb]{revtex4}

\usepackage{graphicx}
\usepackage{dcolumn}
\usepackage{bm}

\begin{document}


\title{Micromagnetic simulations of small arrays\\
of submicron ferromagnetic particles}

\author{Christine C. Dantas}
 \email{ccdantas@iae.cta.br}
\author{Luiz A. de Andrade}
 \email{andrade@iae.cta.br}
\affiliation{%
Divis\~ao de Materiais (AMR), Instituto de Aeron\'autica e Espa\c co (IAE),\\
Comando-Geral de Tecnologia Aeroespacial (CTA), Brazil
}%

\date{\today}

\begin{abstract}
We report the results of a set of simulations of small arrays of submicron ferromagnetic particles. The actions of dipolar and exchange interactions were qualitatively investigated by analysing the ferromagnetic resonance spectra at $9.37$ GHz resulting from the magnetization response of connected and unconnected particles in the array as a function of the applied {\it dc} magnetic field. We find that the magnetization precession movement (at resonance) observed in individual particles in the array presents a distinctive behaviour (an ``amplitude mismatch") in comparison to isolated, one-particle reference simulations, a result that we attribute to the action of interparticle dipolar couplings. Exchange interactions appear to have an important role in modifying the spectra of connected particles, even through a small contact surface.
\end{abstract}

\pacs{75.30.Ds, 76.50.+g, 73.21.-b}


\maketitle

\section{\label{Intro}Introduction}

The main theoretical framework that defines micromagnetism has been established and refined several decades ago by Landau, Lifshitz and Gilbert, among others \cite{LL1935,Gil1955,Bro1963}. Micromagnetism describes magnetic phenomena at scales of $\sim 10^{-6}-10^{-9}$ m and offers a formal basis for the comprehension of emerging and collective phenomena in magnetic materials, as for instance magnetic domains in ferromagnets \cite{Bro1963}, collective excitations of spins  (``spin waves") \cite{Dys1956,Kit1958,Bar2007}, etc.  It is based on a continuum, semi-classical dynamical model for the gyromagnetic precession, including a phenomenological damping term.

During several years \cite{Aha2001}, micromagnetism has been somewhat overlooked  due to difficulties in devising high-resolution (submicrometric) experiments to test it. In addition, the fundamental equation of micromagnetism, known as the Landau-Lifshitz-Gilbert equation (hereon, LLG) \cite{Bro1963}, can only be solved analytically for special cases or supposing considerable simplifications for the description of the problem \cite{dAq2004}. This situation changed rapidly in the past few years, with the development of improved experimental techniques and by the use of numerical simulations with exponentially increasing performance rates, allowing a deeper understanding of the physical phenomena encompassing the realm of magnetism at submicroscopic scales \cite{Fid2000,Mor2007,Gli2008}. These studies mainly aim at the application of microelectronics and data storage \cite{New2004}. In particular, the study of the role of the dynamics of confined spin waves in patterned arrays of magnetic elements in thin films \cite{Jung2002a,Jung2002b,Yu2004,Gub2005,Bai2006} is of great interest in those applications, given that the excitations of spin waves effectively limit the time scale of the magnetization reversal process.

The main purpose of this work is to study the ferromagnetic resonance (hereon, FMR) spectra emerging from small arrays of ferromagnetic particles, with different spatial configurations and geometries, using as references single, isolated one-particle simulations. The idea is to identify qualitatively important contributions from dipolar and exchange couplings between spins as a function of the array configuration.  In order to obtain an overall understanding of these contributions, we have studied configurations varying from particles separated by a small distance, to particles slightly touching each other, or being arbitrarily connected with ferromagnetic material. The present simulations were shown to be computationally demanding, so at this time we have not surveyed the problem through an exhaustive number of array configurations, but instead focused on general trends from some representative cases.

This paper is organized as follows. A theoretical background summary and the time domain micromagnetic simulation setups are given in Sec. I. In Sec. II, we describe the FMR spectra of the array configurations and the equilibrium magnetization fields. In Sec. IV, we discuss the results.

\section{\label{Simul}Micromagnetic Simulations}

\subsection{\label{Theo}Theoretical Background}

Let the magnetization vector $\vec{M}$ be defined as the sum of $N$ individual magnetic moments $\vec{\mu} _j$ ($j=1, \dots , N$) in a small volume $dV$ at a given position $\vec{r}$ of a ferromagnetic particle, namely:  $\vec{M}(\vec{r}) \equiv \frac {\sum_{j=1}^{N} \vec{\mu}_j}{dV}$. Micromagnetism assumes that the direction of $\vec{M}$ varies continuously with position \cite{Bro1963}. The dynamics of the magnetization field $\vec{M}(\vec{r},t)$ under the action of a external magnetic field $\vec{H}_{ext}$ is that of a precession movement of $\vec{M}$ around an effective magnetic field $\vec{H}_{eff}$, defined as $\vec{H}_{eff} \equiv -\mu _0^{-1} \frac{\partial E_{eff}}{\partial \vec{M}}$. In this expression, $E_{eff}$ represents the energy associated with the effective magnetic field, and is given by the sum of four fields representing different interactions among the magnetic moments (spins) of the magnetic material, namely: $E_{eff} = E_{exch} + E_{anis} + E_{mag} + E_{Zee}$. These four terms are, respectively: the exchange energy, the anisotropy energy, the magnetostatic or dipolar energy and the Zeeman energy (namely, the energy associated with the external magnetic field $\vec{H}_{ext}$). The equilibrium state of such a ferromagnetic system is that in which the total energy is minimized.

The Landau-Lifshitz-Gilbert equation mathematically describes the dynamics outlined above, with the following additional physical parameters specified: the saturation magnetization, $M_s$ (determined by the temperature, here fixed throughout), the gyromagnetic ratio, $\gamma$, and a phenomenological damping constant, $\alpha$. The LLG equation is then given by:

\begin{equation}
\frac{d \vec{M}(\vec r,t)}{d t} = 
- \gamma  \vec{M}(\vec r,t) \times \vec{H}_{eff}(\vec r,t) - 
\frac{\gamma \alpha}{M_s} \vec{M}(\vec r,t) \times 
\left [ \vec{M}(\vec r,t) \times \vec{H}_{eff}(\vec r,t) \right ].
\label{LLG}
\end{equation}
Note that the magnetization vector precesses around the $\vec{H}_{eff}$ field and looses energy to the environment in accordance with the damping constant, tending therefore to align with $\vec{H}_{eff}$, given sufficient time span.

The magnetization precession movement can proceed uniformly or not. The latter case leads to spin waves. Their occurence was confirmed experimentally and their behavior has been the object of several numerical investigations and comparative studies with the initial theoretical predictions \cite{Pli1999,Neu2006,Bar2007}. We briefly summarize these possible magnetization precession movements in a small ferromagnet. The uniform precession movement of $\vec{M}(\vec r,t)$ about the direction of an effective local field will occur at a given frequency $\omega_0$. The application of an oscillating magnetic field $\vec{H}_{ac}$ at $\omega_0$, perpendicularly to the former field, will result in a coupling of $\vec{M}$ and $\vec{H}_{ac}$ with energy absorption from the {\it ac} field by the system (leading to an ``uniform resonance" or major peak in the FMR spectrum). But the {\it ac} field may also couple to nonuniform (spin wave) modes of precession of the $\vec{M}$ field. Exchange and dipolar interactions contribute to the energy of these modes. The former (exchange) interactions are expected to be more dominant at physically smaller magnetic elements, leading to additional resonances at frequencies $\omega_p = \omega_0 + D k_p^2$, according to the Kittel's model \cite{Kit1958}, where $D$ is a factor that depends on the exchange  interaction between adjacent spins, and $k_p$ is the (quantized) wave vector corresponding to a given spin wave excitation in the ferromagnet. The resonant peaks associated with the exchange interactions lie at the left of the uniform peak \cite{Jung2002a, Jung2002b}, that is, at smaller values of the {\it dc} component of the applied field. The later (dipolar) interactions, on the other hand, are relatively independent of the size of the ferromagnet, and may be important in a lattice of ferromagnetic elements \cite{Jung2002a}, leading to an interparticle dipolar coupling field.

In order to probe the spin standing modes of coupled dots, two equivalent methods are available. The first one fixes the frequency of an small amplitude applied $ac$ field for different values of a static magnetic field\cite{Jung2002a}. This is the method adopted in the present work (see next section). The second one fixes the static field and probe the frequencies of all the (quasi-uniform and non-uniform) modes. The former method is suitable to compare with FMR results, the latter with Brillouin scattering (BLS) results\cite{Mat1997}. The behavior and character of the modes are the same. The latter method has been developed since about 2000 (pioneering work of Jorzick et al. \cite{Jor2000}) and since then the dipolar-exchange nature and the symmetry of the spin modes of submicrometric dots has been fully understood. In particular, modes with nodal planes either parallel or perpendicular to the static applied field and edge modes have been identified, in addition to the quasi-uniform mode (see Gubbiotti et al.\cite{Gub2005} and references therein).  It is accepted that the modes with nodal planes parallel to the magnetization are high frequency modes (or equivalently on the ``left'' of the FMR peak) and modes with nodal planes perpendicular to the magnetization can exhibit frequencies lower and higher than the quasi-uniform mode (or, in other words, on the ``right'' and ``left'' of the FMR peak), depending on the number of nodes and the balance between the dipolar and exchange effects.

The description of spin waves in a magnetic element can be given in analogy to a vibrating membrane \cite{Bai2006}, but in the magnetic case two additional ``restoring forces" take place, leading to a more complicated description of the normal modes than in the membrane case, where the description is made in terms of sinusoidal standing waves with a unique restoring force acting on the membrane. A thorough review of confined spin waves can be found in Demokritov et al. \cite{Dem2001}.

\subsection{\label{Setup}Simulations Setup}

We have performed micromagnetic simulations based on the numerical integration of the LLG equation (c.f. Eq. \ref{LLG}) using the freely available integrator OOMMF (Object Oriented Micromagnetic Framework) \cite{OOMMF}. In the present work, we have focused our investigations on a set of different small ferromagnetic (permalloy $Ni_{80}Fe_{20}$) circular particle arrays with a small, finite thickness. Full (3D) simulations of the magnetization vector dynamics were performed, and interparticle magnetostatic or dipolar interactions were explicitly considered in the computations. Simulations of isolated, single particles, were also performed for comparison purposes.

\begin{figure}
\includegraphics[scale=0.4]{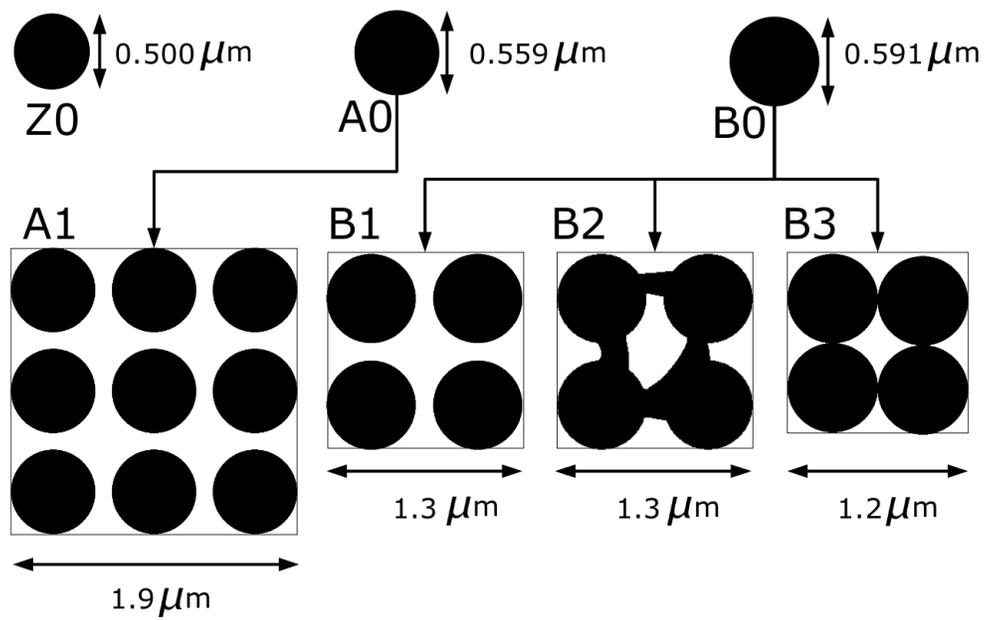}
\caption{\label{fig1} Particle configuration geometries, arranged in ``families'' of common characteristics.}
\end{figure}

The main global parameters of the OOMMF simulator are listed in Tab. \ref{tab-simul}, and were fixed for all simulations here considered. Only parameters dependent on the specification of the particle geometries differ (these can be seen in Fig. \ref{fig1}), as well as the number of external applied {\it dc} magnetic fields used to generate the FMR spectrum in each simulation.  Table \ref{tab-geom} lists specific data of the simulations that may differ from each other, e.g.: the diameter of the particles ($d$), the diameter of the array ($D$), the interparticle spacing ($a$; i.e., the distance between the centers of two adjacent particles, or lattice spacing) and other simulation parameters, to be discussed in a moment.  

The arrays hold a small number of particles each, arranged in a regular, $2 \times 2$ (four particles, B-``family'') or $3 \times 3$ (nine particles, A-``family''), grid (2D square lattice); see Fig. \ref{fig1}. In the case of four particle arrays, we have studied three types of configurations: equally spaced particles (B1 configuration), equally spaced particles with arbitrary connections among them (B2; in this case we define the interparticle spacing $a$ as that of B1, although the particles are connected among themselves)\footnote{Our motivation for considering the B2 configuration was the study of the relative contribution of dipolar and exchange interactions for complex geometries. In addition, as one can observe in the paper of Jung et al. \cite{Jung2002a} (their Fig. 1), the patterning procedure appears not to be perfect and often produces artifacts between the dots. We have attempted to include arbitrary ``bridges'' between elements, with no particular criteria for their geometrical shape. Finally, the understanding of the micromagnetics of such complex geometries is being currently conducted for specific applications of interest at the AMR/IAE/CTA Division.}, and particles ``touching'' each other (B3). As a reference, we have considered isolated particles (labelled Z0, A0 and B0), which differ slightly in diameter, according to the particle diameters of the corresponding configurations, as indicated in Fig. \ref{fig1}. Notice that we have generally prioritised the defintion of the array diameter with the choice of more ``rounded''  values for the discretization (defined by the ``cell size'' parameter; see Tab. \ref{tab-geom}) over individual particle diameters, but such a choice is immaterial; the effect of different relative particle diameters are considered in the analysis. The Z0 particle simulation was included in order to compare with previous work by Jung et al. \cite{Jung2002a,Jung2002b}.  

Larger arrays of particles were not considered at the present time, given the high computational demand of these simulations. Indeed, in order to obtain accurate results, the value of the cell size should not exceed the exchange length \cite{dAq2004}, defined as $l_{ex}=\sqrt{2A/(\mu_0 M_s^2)}$, which in the present work results in $l_{ex} \sim 5.7$ nm. Most of the simulations were executed on a 3 GHz Intel Pentium PC running Kurumin Linux and on a 2GHz Intel Core Duo running Mac OS X, and each simulation took between $\sim 6$ to $28$ hours of CPU time, depending on the configuration.

\begin{figure}
\includegraphics[scale=0.5]{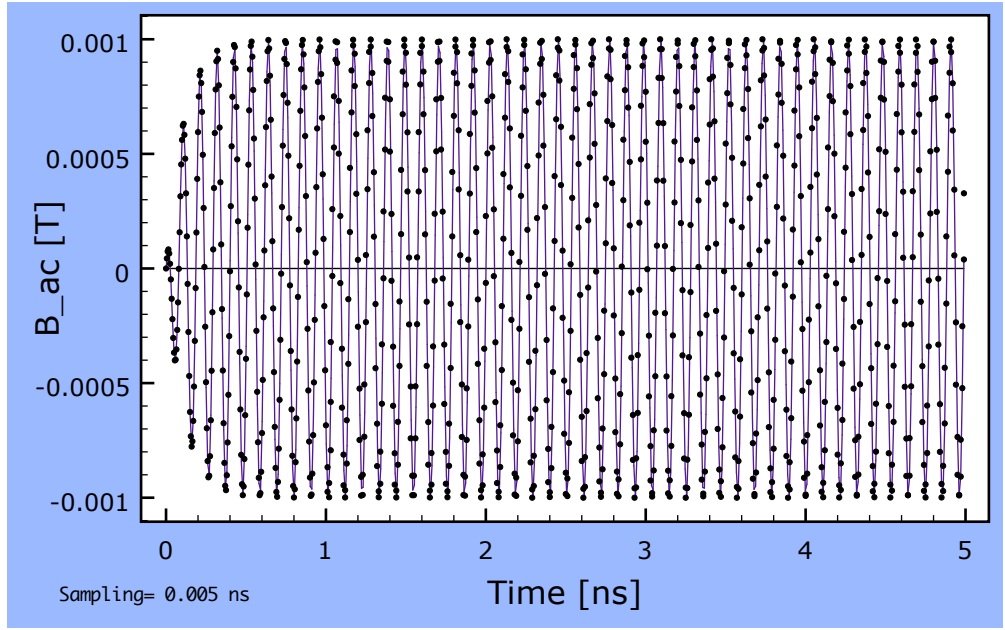}
\caption{\label{fig2} The time dependence and discretization of the applied $\vec{B}_{ac}$ field.}
\end{figure}

In order to obtain the ferromagnetic resonance of each of the configurations previously described, the following prescription was adopted. An external magnetic field in the plane of the particles was applied, formed by two components: a static ({\it dc}) magnetic field ($\vec{B}_{dc} \equiv \mu_0\vec{H}_{dc}$) in the $y$ direction, and a varying ({\it ac}) magnetic field ($\vec{B}_{ac} \equiv \mu_0 \vec{H}_{ac}$) of small amplitude in the $x$ direction, conforming with Jung et al. \cite{Jung2002b}:
\begin{equation}
\vec{B}_{ac} = (1-e^{-\lambda t}) \vec{B}_{ac,0} \cos (\omega t),
\end{equation}
with the {\it ac} field frequency given by $f = \omega / (2 \pi) = 9.37$ GHz, $\lambda \sim f$, and $\vec{B}_{ac,0}=1$ mT. Fig. \ref{fig2} shows the time dependence and discretization of the applied $\vec{B}_{ac}$ field. The values of the $\vec{B}_{ac}$ field at intervals of $0.005$ ns were used as inputs in the ``field range'' record of OOMMF, which in turn were stepped linearly by the simulator. The simulations were run up to $5$ ns, resulting in $1000$ outputs for each simulation (i.e., one simulation for each value of the $\vec{B}_{dc}$ field), as indicated in Fig. \ref{fig2}. We have performed a set of simulations for each array configuration by varying the range of the $\vec{B}_{dc}$ field from $0.00$ T to $0.39$ T, at intervals of $0.01$ T, resulting typically in $40$ simulations for each configuration, as indicated in the last column of Tab. \ref{tab-geom}. 
The saturated regime is at $\vec{B}_{dc} > 0.5$ T and therefore outside our analysis. Some configurations present a few more than $40$ simulations (B0, B1); in these cases, the additional simulations were performed more fine-grainly around the resonance peak in order to evaluate whether significant divergence was found in the results (see next section). The A1 configuration had a fewer number of simulations due to high computational demands; hence a more coarse-grained ``sampling'' of the underlying FMR spectrum was obtained in this case, as compared to the other configurations. Finally, we point out that the magnetization field of the particles was initially aligned to the same direction of the external $\vec{B}_{dc}$ field. 

\section{\label{Resul}Results}

In Fig. \ref{fig3}, we present the simulated time dependence of the spatially averaged magnetization vector $\vec{M}$, in the $x$ direction, normalized by the saturation magnetization ($M_s$). The various curves represent the resulting time dependence for each array/particle simulation evaluated at their respective FMR peak outputs, to be discussed below. The curves were offset for clarity. The A0 simulation is not shown due to the fact of being very similar to the B0 one, so it is omitted for clarity. The time dependence of the $\vec{B}_{ac}$ field (arbitrarily normalized in the main figure and in the insets) is also included for comparison purposes. The insets in Fig. \ref{fig3} correspond to zoom-in regions of the initial time evolution (left inset) and the steady state regime (right inset). Notice that all magnetization field responses are out of phase with the applied field after transient effects vanish, and the phase responses are practically identical in all cases, only differing in amplitude. Interestingly, the B3 configuration starts off at a different phase, but catchs up soon after the first cycle. From that figure it is already clear that the magnetization field of all array simulations have, on average, a smaller response to the external field than the corresponding reference, one-particle simulations (represented by the B0 simulation in the figure).

In order to obtain the FMR spectra of the configurations, we proceeded as follows. The first $3$ ns of all data have been excluded. For each simulation in a given configuration (i.e., for each applied {\it dc} magnetic field), the Fourier transform of the spatially averaged magnetization vector $\vec{M}$, in the $x$ direction, was obtained (as already mentioned, Fig. \ref{fig3} refers to the results at the FMR peak only). The amplitude of the maximum Fourier peak at each $\vec{B}_{dc}$ field was then obtained, resulting in the FMR spectra of Fig. \ref{fig4}. In Fig. \ref{fig5}, we show the derivatives of the FMR spectra. Regarding the main body of Fig. \ref{fig4}, we have applied a spline fit to the reference single particle simulations' data (Z0, A0, B0), but maintained the individual data points of the array simulations. In the insets of Fig. \ref{fig4}, however, all data has been spline fit to facilitate the comparison of the overall behavior of the curves. The derivative FMR spectra (Fig. \ref{fig5}) were obtained by derivation (in intervals of $0.022$ T) of the latter splined curves.

We report the following three overall observations. First, the resonance uniform mode peaks of the reference single particle simulations show the expected trend \cite{Jung2002b}, namely, a shift in the peak position as a function of the diameter of the particle (the peak position shifts towards smaller values of the external field $\vec{B}_{dc}$ as the particle diameter increases). The trend is very small, given that the particles differ only slightly in diameter. Second, {\it (i)} the resonant uniform mode peak position of the B1 array configuration is shifted towards smaller values of the external field $\vec{B}_{dc}$ as compared to the reference (B0) one-particle simulation, even though the particles that compose the B1 array have the same diameter of the B0 particle. Also,  as already observed in Fig. \ref{fig3}, {\it (ii)} the amplitude of the uniform mode peak of the B1 configuration is smaller than that of B0. Finally, {\it (iii)} the secondary peak at the right of the uniform mode peak in B0 (also clear in Z0 and A0, see arrows in Fig. \ref{fig5}) is no longer apparent in B1. These statements, namely, {\it (i)}-{\it (iii)}, are also applicable to the A1 configuration, although the effects seem to be slightly more pronouced in this case, as one can notice by inspecting the derivative FMR spectra of Fig. \ref{fig5} (see the dashed lines connecting the corresponding resonance peaks).  Third, as the particles of the B1 array are arbitrarily connected (B2) or approach to the point of ``touching'' each other (B3), the resulting spectra significantly evolve away from the B0-B1 spectra: {\it (a)} the amplitude of the uniform mode peak of both B2 and B3 decrease and the overall spectrum gets more spread, with the presence of secondary contributions at the left of the uniform peak position (B2 and B3), as well as to the right of it (in the case of B2). {\it (b)} The B2 uniform mode peak does not appear to shift in position relatively to the B1 peak, whereas the B3 does, towards {\it  higher} values of the $\vec{B}_{dc}$ field. 

In order to better understand the observed characteristics of the FMR spectra, we analysed in more detail the aspect of the magnetization fields at the uniform resonance peak. In Fig. \ref{fig6} the ``snapshots'' of the magnetization vector field (at FMR) at four points of the cycle ($\omega t = 0, \pi/2, \pi, 3\pi/2$) are shown, being selected  from a cycle around $\sim 4$ ns (when transient effects are over). Different pixel tonalities correspond to different values of the $x$ component of the magnetization vector field, which in turn was subsampled to display an arrow for the average of $9$ vectors per cell element. Notice that we have included a sinusoidal function at the top marking the four points corresponding to the selected ``snapshots'', so that immediately below each point of the reference curve the magnetization state at the cycle point can be directly observed and compared with that of the reference single particle simulation. 

The most obvious feature of Fig. \ref{fig6} is that the individual particles composing the array configurations do not oscillate in synchrony with the corresponding single particle simulations. When averaging out the magnetization field over the array particles, the emerging response is lower than that of single particles, resulting in the observed lower amplitude of the uniform peak resonance of the spectra (Figs. \ref{fig4} and \ref{fig5}). Indeed, one could naturally expect that the individual particles in the array configurations B2 and B3 would have their magnetization field evolving somewhat differently than that of the single particle simulations, since these configurations physically join the particles of the array, removing the circular geometry characterization of the individual particles (forming effectively a ``larger" single particle with a different geometry). But in the case of the A1 and B1 configurations, the most suspicious agent causing the desynchronization would be an interparticle dipolar coupling field.

In order to illustrate more clearly the behavior of the oscillating magnetization fields, we have synthetized the main information of the oscillating pattern into a ``correspondence scheme". We have used the single particle simulations to associate a circular symbol of a diameter proportional to a given cycle point aspect (the pixel tonalities distribution) of the magnetization field. Fig. \ref{fig7} illustrates the correspondence scheme adopted.  We have used this correspondence scheme to recast the previous results (Fig. \ref{fig6}) into the diagrammatical forms of Fig. \ref{fig8} (excluding the B2 configuration, which cannot be represented in a simple manner in the scheme). Under this new representation, the underlying oscillatory patterns are more clearly displayed or synthetized. One can see that the suspected desynchronization is an ``amplitude mismatch'': {\it the amplitude of the x-component of the magnetization as a function of position in the magnet, in each of the particles in a given array, is not similarly distributed among the particles}. 

Another interesting point is the following. If one recast the above representation into a matrix representation (B configurations as $2 \times 2$ matrices and the A1 configuration as a $3 \times 3$ matrix), one can observe that the $B_{1,j}$ elements behave in opposition to the $B_{2,j}$ ones ($j = 1, 2$). In the case of the A1 matrix, one can notice a similar trend: the elements in the $A1_{1,j}$ and $A1_{3,j}$ rows ($j = 1, ..., 3$) evolve in opposition to each other. The central row $A1_{2,j}$ ($j = 1, ..., 3$) appears to evolve independently.

\section{\label{Disc}Discussion}

The simulations of small arrays of ferromagnetic particles here performed indicate that their spatial configurations influence the resulting FMR spectra in distinctive ways. Here we discuss the possible effects of these spectra from the point of view two physical interactions: dipolar and exchange mode interactions.

\begin{figure}
\includegraphics[scale=0.5]{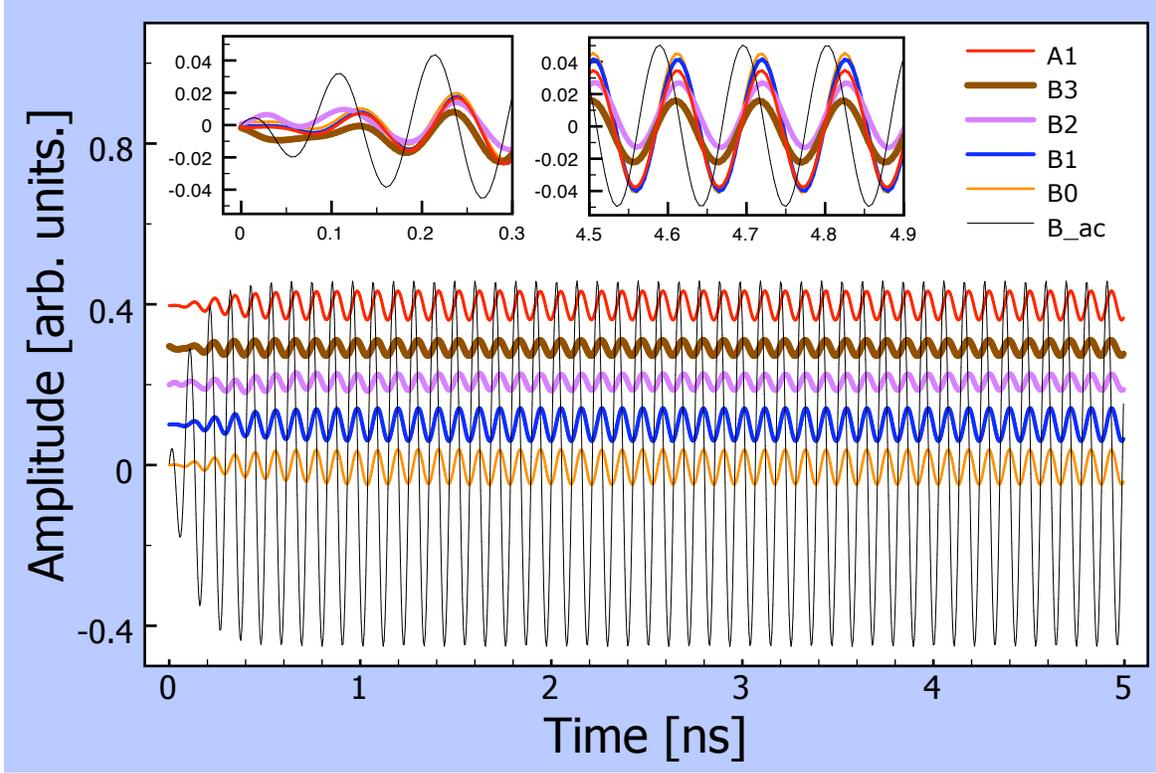}
\caption{\label{fig3} Time dependence of the amplitude (at FMR) of the spatially averaged magnetization vector $\vec{M}_x/M_s$ of the array configurations. In the main body of the figure, the curves were offset for clarity, and are shown at the same presentation order as the legend (top curve is that of the A1 configuration). The time dependence of the external $\vec{B}_{ac}$ field (arbitrarily normalized; thin line) is included.}
\end{figure}

\begin{itemize}
\item {Dipolar interactions between the particles:}
\subitem {{\it A1 and B1 configurations:} since in these cases the particles are not physically in direct contact, we may presume that the dipolar interaction is causing the moderate decrease in amplitude of the uniform mode peak in comparison to the one-particle reference simulations. Indeed, one can suppose that the finiteness and symmetry of the arrays over which the particles are distributed favor an interparticle dipolar interaction for which the combined effect is that of a mismatching of the local amplitudes attained by the  magnetization field precession movement (observed at the uniform mode outputs).  Another observation is the shift of the uniform mode peak towards a smaller value of the applied dc magnetic field, in comparison to the one-particle reference simulations. It mimmics the expected spectrum of an effective single particle of larger diameter. One may notice that in the case of the A1 configuration the {\it whole} spectrum seems to shift accordingly (see dashed lines of Fig. \ref{fig5}); this does not seem to be the case of the whole B1 spectrum. The effect, clearly seen in the A1 configuration, has already been experimentally measured with BLS and calculated by Gubbiotti et al. \cite{Gub2006} with the second alternative method mentioned in Sec. II.A for a similar $3 \times 3$ array. The shift towards lower values of the static field of the modes corresponds to the increase of the mode frequencies in Fig. 3 of Gubbiotti et al. A direct comparison between the modes found in the simulation of Gubbiotti et al. and the ones in the present work is difficult because in the former case the dynamic magnetization is shown, while in this work (c.f. Fig. \ref{fig6}) the total magnetization (static plus dynamic) is shown. We intend to proceed these investigations in a future work. Regarding the nonuniform modes of precession, one can observe the following facts. First, the lowest energy mode (at the right of the uniform mode), seen in the reference-one particle simulations (see arrows in Fig. \ref{fig5} over the derivative spectra of Z0, A0 and B0), diminish for the A1 and B1 configurations (one cannot affirm that they completely vanish, given the smothing over a limited resolution in $\vec{B}_{dc}$ used to derive the spectra). Second, a relatively larger gain in response of the nonuniform modes at the left of the uniform mode peak is found (c. f. main body of Fig \ref{fig5}). This will be discussed in more detail below, since these peaks are thought to arise from exchange interactions, although will possible contributions from dipole interactions.}
\subitem {{\it B2 and B3 configurations:} In these cases, the particles are directly in contact. In the case of B3 configuration, the ``amplitude mismatch'' effect seen in the previous cases is again clearly observed (see Figs. \ref{fig6} and \ref{fig8}). Yet, from these figures one can qualitatively observe that, looking at individual particles in the array, the magnetization field precess in smaller amplitude as compared to the B0, B1 configurations, although in similar pattern as the latter, resulting in a smaller relative amplitudes of the average magnetization field, which one can confirm from an inspection of Fig. \ref{fig3}. Although we have not traced an ocillatory pattern scheme for the B2 configuration, it is remarkable that it somewhat seems to follow the pattern observed in the B1 configuration (as one can observe looking only at the central parts of the particles of the B2 array, excluding the dynamics occuring at their connections, c.f. Fig. \ref{fig6}), from which the latter differs from the former only by the application of arbitrary connections. The amplitude of the average magnetization field of the B2 configuration is greater than that of the B3 configuration, possibly because the contribution coming from the central region of the particles of the B2 configuration can follow somewhat that of the B1 ones, yet being significatively smaller than the latter ones because of the averaging over the field behavior of the connected parts. Evidently, the resulting spectrum will be more complex in this case, with, for instance, the presence of a new nonuniform peak at the far right of the uniform one (see the arrow over the B2 curve of Fig. \ref{fig5}). Another distinct aspect of the B3 configuration is the shift of the uniform mode peak towards a {\it larger} value of the applied {\it dc} magnetic field, in comparison to the one particle reference simulation, contrary to the trends of the A1 and B1, B2 configurations, in which the peaks move to the opposite direction. The origin of the B3 shift is also unclear at this point and deserves further investigation.}
\item {Exchange mode interactions:}
\subitem {{\it A1 and B1 configurations}: In these cases, the first nonuniform peak at the left of the uniform one does not appear to suffer significant changes in amplitude as compared to the reference one-particle simulations. However, as already mentioned, the spectrum of the A1 configuration is clearly shifted as a whole to the left. The other higher harmonics to the left of the uniform mode from both A1 and B1 seem to have slightly higher amplitudes in comparison to the one-particle reference simulation (spectra between $\vec{B}_{dc}=0, ..., 0.08$ T). Since the particles in these configurations are not in physical contact, the energy of these nonuniform peaks, understood to be mainly the result of exchange mode interactions, must have been suplemented by the injection coming from other means. In this case,  dipolar interaction is a reasonable candidate, since the particles do not touch each other. This assertion would agree with the hypothesis raised by Jung et al. \cite{Jung2002a} on the possible coupling of the dipolar interactions with exchange spin wave modes.}
\subitem {{\it B2 and B3 configurations:} In these cases, as expected, the spectra evolve in a complex way in comparison to the reference simulations. In particular, one can observe a relatively large gain in response of the nonuniform modes (Fig \ref{fig5}). As observed previously, a fraction of this gain should be a result of the coupling of the dipolar and exchange interactions. But since the particles of these arrays are in physical contact, the exchange interactions should have a more important role. One remarkable aspect is the form of the B3 configuration spectrum which is quite different from that of the B1 configuration, and in some aspects approach more closely to that of the B2 configuration (c.f. Fig. \ref{fig5}). This fact implies that propagative effects arising from exchange interactions, even when are able to act through a small contact surface, can result in significant modifications in the FMR spectrum.}
\end{itemize}

\begin{figure}
\includegraphics[scale=0.6]{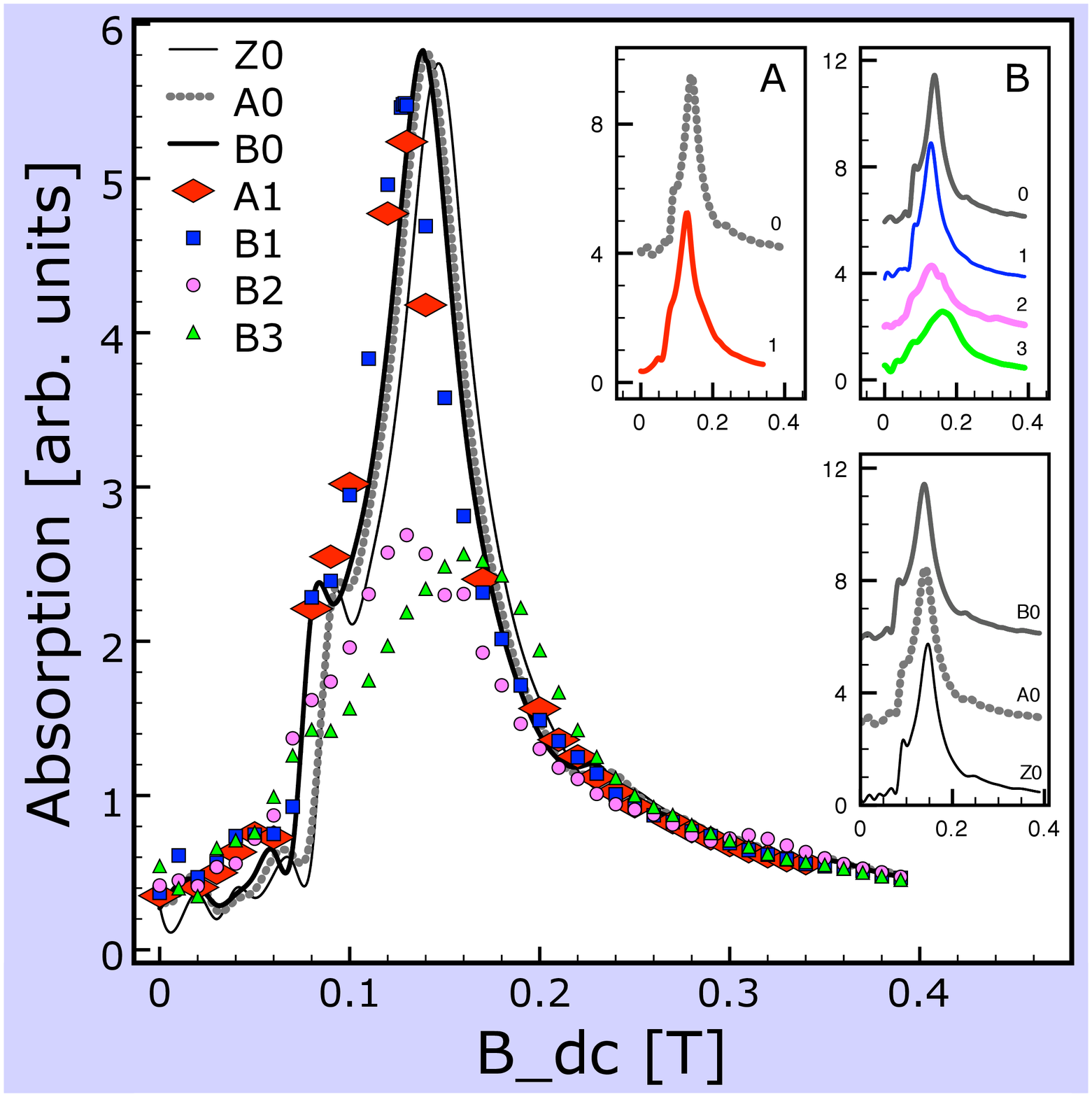}
\caption{\label{fig4} The ferromagnetic spectra of all configurations.}
\end{figure}

\begin{figure}
\includegraphics[scale=0.6]{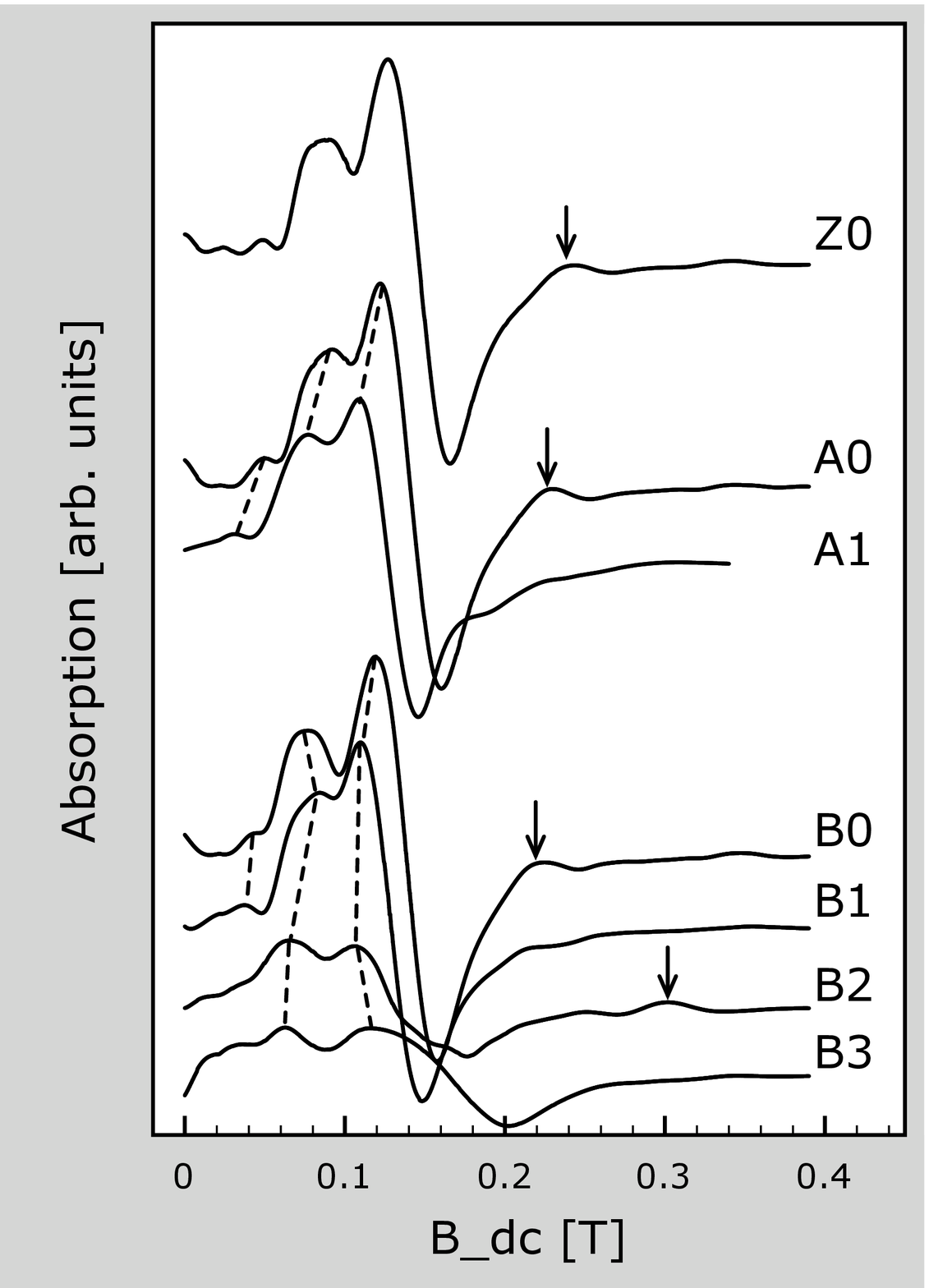}
\caption{\label{fig5} The derivative ferromagnetic spectra of all configurations.}
\end{figure}

\begin{figure}
\includegraphics[scale=0.5]{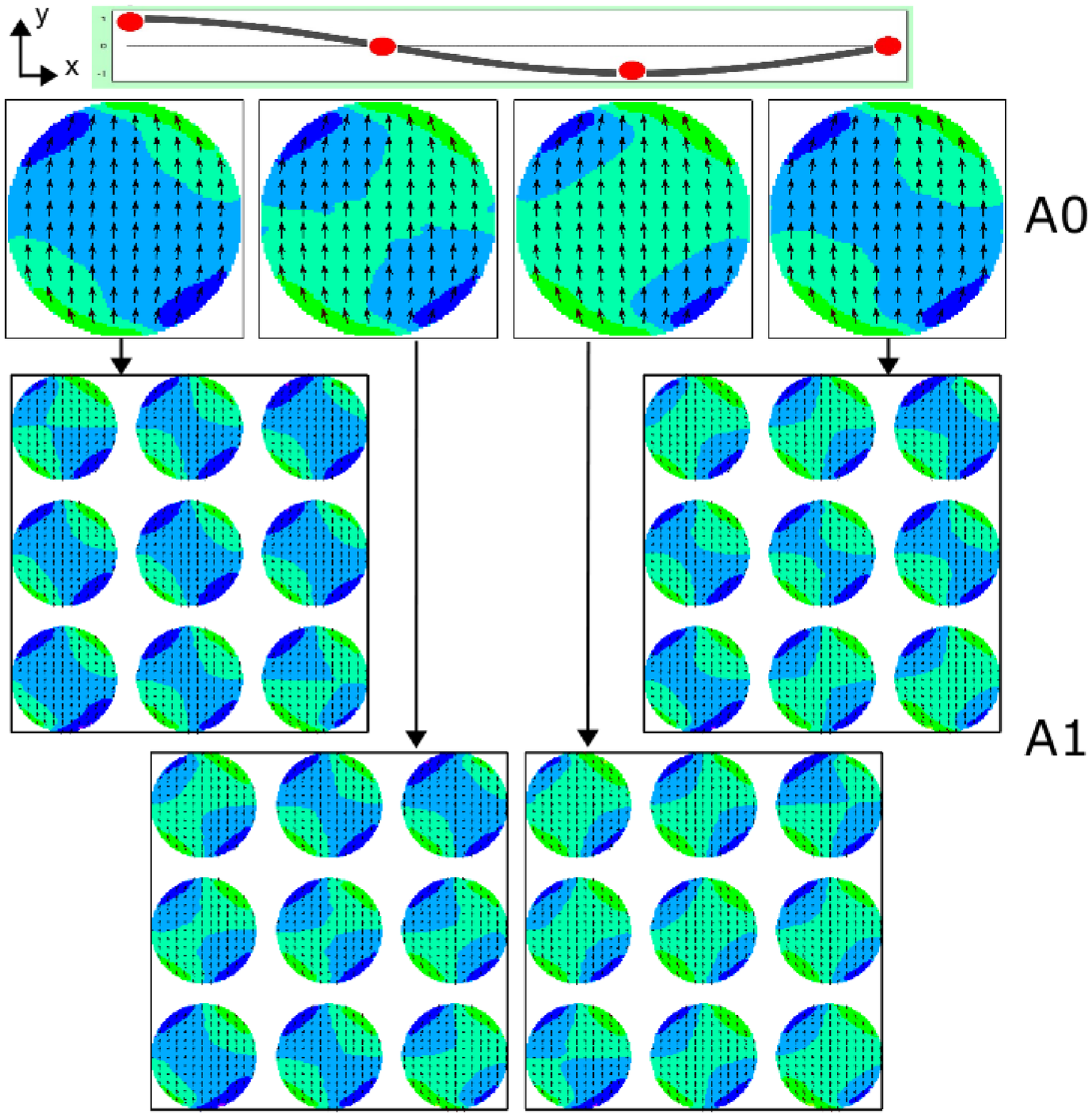}
\includegraphics[scale=0.5]{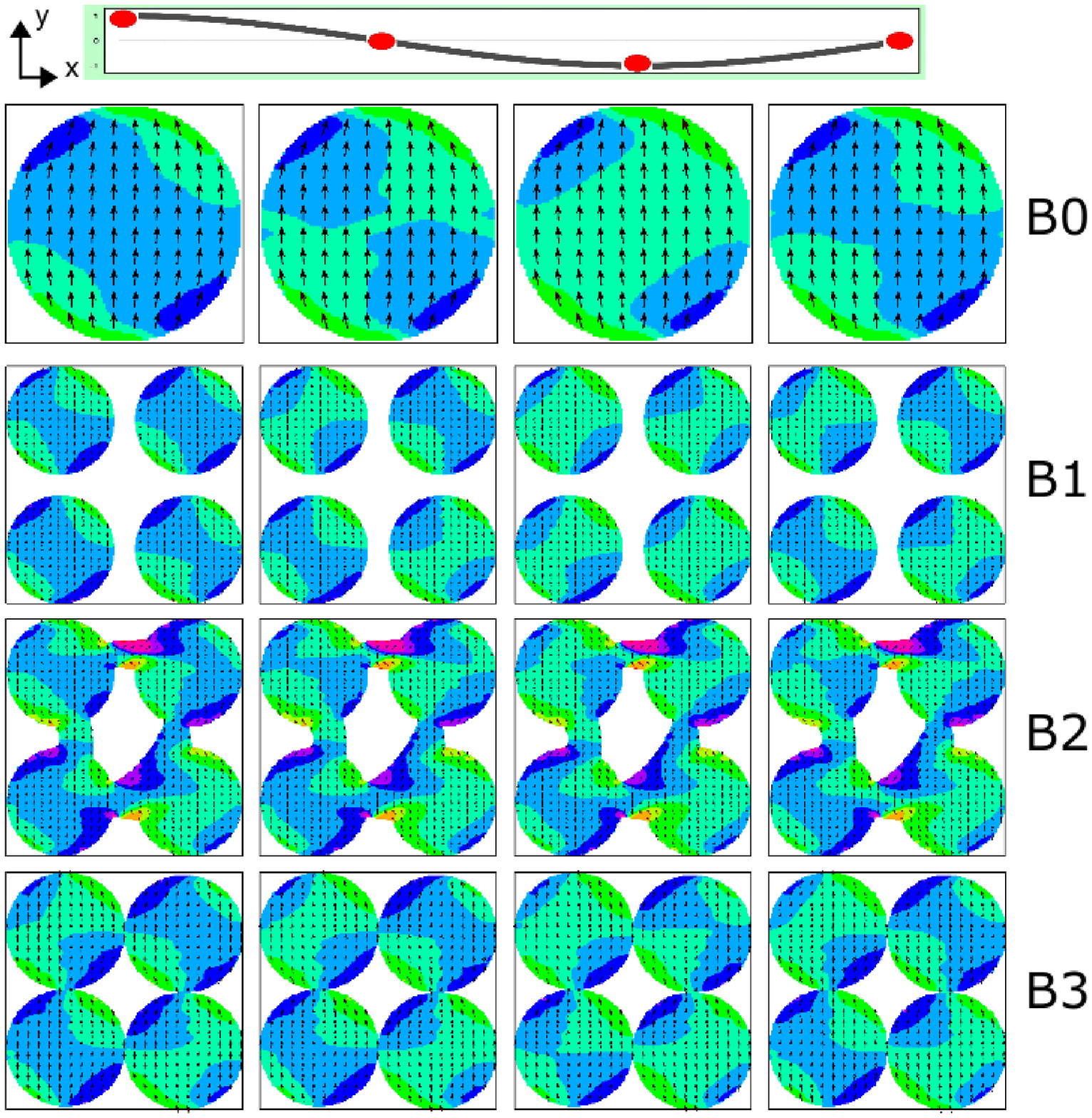}
\caption{\label{fig6}  ``Snapshots'' of the magnetization vector field (at FMR) at four points of the cycle ($\omega t = 0, \pi/2, \pi, 3\pi/2$; see sinusoidal curve at the top of the snapshots) after $\sim 4$ ns (when transient effects are over). Different pixel tonalities correspond to the value of the $x$ component of the magnetization vector field at the pixel element. }
\end{figure}

\begin{figure}
\includegraphics[scale=0.5]{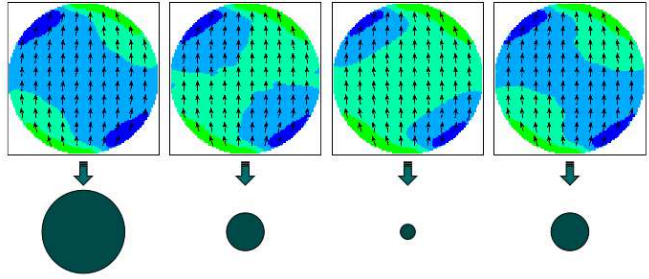}
\caption{\label{fig7} The oscillatory pattern correspondence scheme (see the main text form further details).}
\end{figure}

\begin{figure}
\includegraphics[scale=0.6]{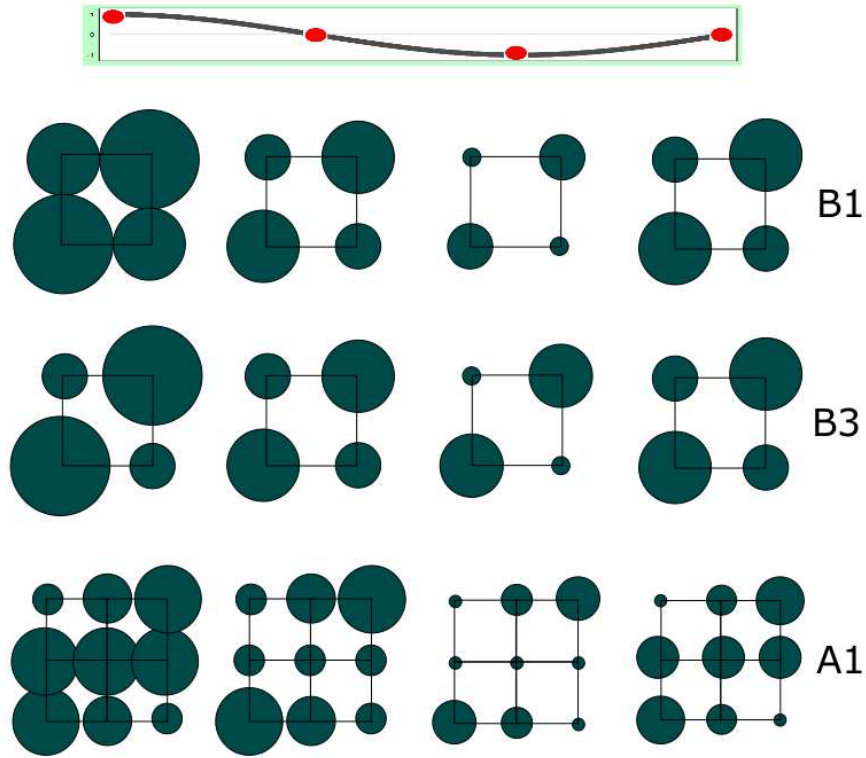}
\caption{\label{fig8} Same as Fig. 6, recasted according to the scheme of Fig. 7, synthetizing the combined effect of the local amplitude of the x-component of the magnetization field at each cycle point of the {\it ac} driving field, for configurations B1, B3 and A1.}
\end{figure}

Finally, an important remark should be mentioned. According to the work of Gubbiotti et al. \cite{Gub2006}, the effect of the interparticle dipolar coupling is to split the modes, spreading them into bands, in the limit of large arrays. In the case of a $3 \times 3$ array, each mode should give rise to $9$ modes, including degeneracy.  Gubbiotti et al. have shown that this broadening is appreciable, even large, for the quasi-uniform mode. However, this effect is not visible in the present study (c.f. cases A0-A1 and B0-B1; Figs. \ref{fig4} and \ref{fig5}). The reasons for the lack of splitting are currently under investigation.

In summary, in the present paper we have presented a set of 3D simulations of small arrays of ferromagnetic particles supposed here to represent small isolated sections of a patterned thin film. We have analysed the resulting FMR spectra and the magnetization field behavior at the resonance modes. We show that the spatial configurations and geometries of the particles in the arrays influence the resulting FMR spectra in distinctive and perhaps unanticipated ways. We have attempted to isolate the action of dipolar and exchange interactions by studying arrays with particles both connected and not connected among themselves. These interactions appear to have an interesting role on the dynamics of the magnetization precession among the particles in the array, as described in detail in this paper (synthetized in Fig. \ref{fig8}). Other simulations are intended to be performed in a future work. In particular, it would be interesting to perform simulations with the applied field at different nominal angles (specially, at $45^{\circ}$), using the same arrays here analysed. According to measurements of Jung et al. \cite {Jung2002a}, the typical low energy peak of one-particle simulations appear more pronounced in these cases. Since this low energy peak is suspected to arise mainly from dipolar interactions, it would serve as an interesting tracer of the role of these interactions in the array.

\begin{acknowledgments}
We would like to thank the attention and technical support of Dr. Michael J. Donahue in the initial phases of this project. We also wish to acknowledge the support and encouragement of Dr. Mirabel C. Rezende throughout this work. Finally, we thank the referees for useful comments and corrections, which helped to improve the paper considerably.
\end{acknowledgments}

\clearpage

\begin{table}
\caption{\label{tab-simul} Main global parameters adopted for the OOMMF simulator, fixed for all simulations in the present work, except for the ``particle width/height'' and ``cell size'' parameters, which are listed separately in Tab. II, as indicated.}
\begin{tabular}{lll}
 Simulation Parameter/Option   & \hspace{1cm} & Parameter Value/Option \\ \hline
Saturation magnetization [A/m] &  & $8.0 \times 10^5$\\
Exchange stiffness [J/m]       &  & $1.3 \times 10^{-11}$ \\
Anisotropy constant [J/m$^3$]  &  & $0.0$ \\
Damping constant               &  & $0.05$\footnote{The value of the damping constant here adopted is far larger than the real one for Permalloy ($\leq 0.01$). A small value of damping constant would allow a better resolution of the absorption lines within the FMR spectra but would lead to prohibitive computation times.}\\
Gyromagnetic ratio [m/(A.s)]   &  & $2.21 \times 10^5$\\
Particle thickness [nm]        &  & $85.0$ \\
Particle width/height          &  & see $D$ in {\sc Table II} \\
Cell size                      &  & see {\sc Table II} \\
Demagnetization algorithm type &  & magnetization constant in each cell\\
\end{tabular}
\end{table}

\clearpage

\begin{table}
\caption{\label{tab-geom} Particular parameter values of the simulations: the diameter of the particles ($d$), the diameter of the array ($D$; for individual particles, $D=d$.), the interparticle spacing ($a$, when applicable), and cell size. Other data are: the number of ferromagnetic particles (column $2$) and the number of external applied {\it dc} magnetic fields (or number of simulations) used to generate the FMR spectrum (column $7$). }
\begin{ruledtabular}
\begin{tabular}{ccccccc}
Label & \# FM particles & $d$ ($\mu m$) & $D$ ($\mu m$) &
 $a$ ($\mu m$) & Cell size (nm) & $\vec{B}_{dc}$ (number of simulations)\\ \hline
Z0 & $1$ & $0.500$ & $0.500$ & --      & $5.00$ & $40$ \\
A0 & $1$ & $0.559$ & $0.559$ & --      & $5.59$ & $40$ \\
B0 & $1$ & $0.591$ & $0.591$ & --      & $5.91$ & $46$ \\
A1 & $9$ & $0.559$ & $1.900$ & $0.671$ & $5.00$ & $26$ \\
B1 & $4$ & $0.591$ & $1.300$ & $0.709$ & $5.00$ & $43$ \\
B2 & $4$ & $0.591$ & $1.300$ & $0.709$ & $5.00$ & $40$ \\
B3 & $4$ & $0.591$ & $1.200$ & $0.591$ & $6.00$ & $40$ \\
\end{tabular}
\end{ruledtabular}
\end{table}

\clearpage

\clearpage

\clearpage

\clearpage

\clearpage

\clearpage

\clearpage

\clearpage

\bibliography{DantasEtAl2008}

\end{document}